\documentclass[
    aps,
    prl,
    reprint,
    nolongbibliography,
    nobibnotes,
    nofootinbib,
]{revtex4-2}

\usepackage[dvipsnames, usenames]{xcolor}
\definecolor{linkcolor}{rgb}{0.1, 0.5, 0.7}

\usepackage[
colorlinks=true,
linkcolor=linkcolor,
citecolor=linkcolor,
filecolor=linkcolor,
urlcolor=linkcolor,
pdfusetitle,
]{hyperref}

\usepackage{amssymb}
\usepackage{amsmath}
\usepackage{physics}
\usepackage{orcidlink}
\usepackage{xspace}
\usepackage{multirow}
\usepackage[absolute,overlay]{textpos}

\newcommand{\chieff}{\ensuremath \chi_\mathrm{eff}\xspace}
\newcommand{\chip}{\ensuremath \chi_\mathrm{p}\xspace}
\newcommand{\mstarcred}{\ensuremath 46.2_{-7.2}^{+12.6}\xspace}

\newcommand{\ligo}{\affiliation{LIGO Laboratory, Massachusetts Institute of Technology, Cambridge, MA 02139, USA}}
\newcommand{\mki}{\affiliation{Kavli Institute for Astrophysics and Space Research, Massachusetts Institute of Technology, Cambridge, MA 02139, USA}}
\renewcommand{\mit}{\affiliation{Department of Physics, Massachusetts Institute of Technology, Cambridge, MA 02139, USA}}
\newcommand{\adler}{\affiliation{Adler Planetarium,1300 South DuSable Lake Shore Drive, Chicago, IL, 60605, USA}}
\newcommand{\ciera}{\affiliation{Center for Interdisciplinary Explorationand Research in Astrophysics (CIERA), Northwestern University, 1800 Sherman Avenue, Evanston, IL, 60201, USA}}
\newcommand{\skai}{\affiliation{NSF-Simons AI Institute for the Sky (SkAI), 172 East Chestnut Street, Chicago, IL, 60611, USA}}
\newcommand{\williams}{\affiliation{Williams College, Williamstown, MA 01267, USA}}

\begin{document}

\begin{textblock*}{4cm}(0.5pt,5pt)
    \includegraphics[width=4cm]{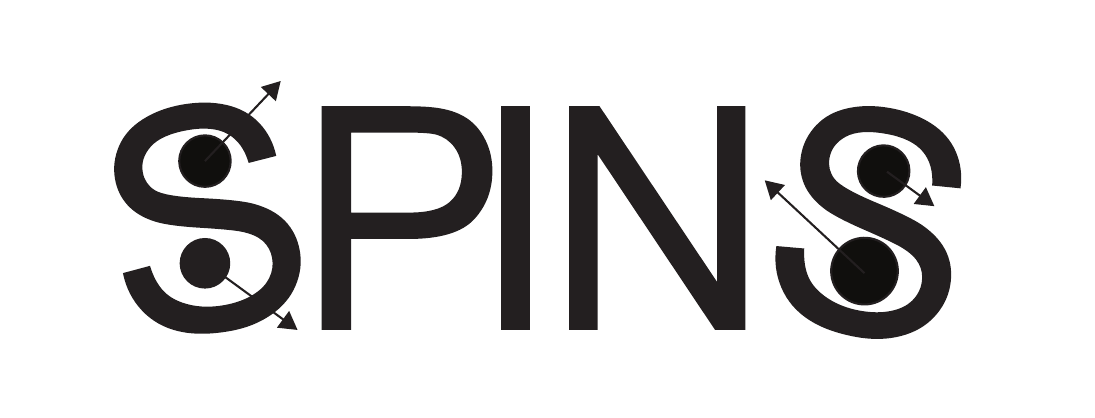}
\end{textblock*}

\title{Signatures of a subpopulation of hierarchical mergers in the GWTC-4 gravitational-wave dataset}

\author{Cailin Plunkett\,\orcidlink{0000-0002-1144-6708}}\email{caplunk@mit.edu}\ligo\mki\mit
\author{Thomas Callister\,\orcidlink{0000-0001-9892-177X}}\williams
\author{Michael Zevin\,\orcidlink{0000-0002-0147-0835}}\adler\ciera\skai
\author{Salvatore Vitale\,\orcidlink{0000-0003-2700-0767}}\ligo\mki\mit
\collaboration{Society of Physicists Interested in Non-aligned Spins, SPINS}
\thanks{\url{www.sites.mit.edu/spins}}
%\date{\today}

\begin{abstract}

Repeated black-hole mergers in dense stellar clusters are a plausible mechanism to populate the predicted gap in black hole masses due to the pair-instability supernova process. These hierarchical mergers carry distinct spin characteristics relative to first-generation black holes. We introduce an astrophysically motivated model in the joint space of effective inspiral and precessing spins, which captures the dominant spin dynamics expected for hierarchical mergers. We find decisive evidence for a transition at $m_1 = \mstarcred M_\odot$, above which the population is nearly entirely hierarchical, a location consistent with the anticipated onset of the pair-instability gap. We also infer a global peak in the hierarchical merger rate at $m_1 = 15.7_{-1.1}^{+3.2}M_\odot$. The existence of low- and high-mass subpopulations of higher-generation black holes suggests the contribution of both near-solar-metallicity and metal-poor star clusters to the hierarchical merger population. Our results reinforce the growing evidence for detailed, mass-dependent substructure in the spin distribution of the binary black hole population.

\end{abstract}

\maketitle

\paragraph*{Introduction}---
\label{sec:intro}
Theoretical models for binary black hole (BBH) formation can be broadly divided into two classes: isolated evolution of field binary stars and dynamical channels~\cite[see, e.g.,][for reviews]{Mandel:2018hfr, Mapelli:2021taw, Mandel:2021smh}.
Identifying the formation history of an individual merger can be difficult since multiple channels may be consistent with a given set of black-hole parameters.
The latest catalog of gravitational-wave (GW) events~\cite{gwtc4cat, gwtc4pop} released by the LIGO--Virgo--KAGRA (LVK) Collaboration~\cite{LIGOScientific:2014pky, Capote:2024rmo, LIGO:2024kkz, VIRGO:2014yos, KAGRA:2020tym} allows for the most detailed investigation yet of the constitution of the BBH population.

It is increasingly clear, both from individual events and population analyses, that massive black holes exist in the range $45\text{--}120\,M_\odot$~\cite{LIGOScientific:2020iuh, gwtc3pop,  Edelman:2021fik, LIGOScientific:2025rsn}. This is in tension with the notion of pair-instability supernovae, which stellar evolution theory predicts should leave no black holes in this mass interval~\cite{Fowler:1964zz, Barkat:1967zz, Rakavy1967, fraley1968supernovae, Heger:2002by, Farmer:2019jed, Stevenson:2019rcw, Farmer:2020xne, Woosley:2021xba}. These observations have spurred further investigation into mechanisms that can populate this gap~\cite{Farrell:2020zju, Baibhav:2020xdf, Tagawa:2020qll, Vaccaro:2023cwr, Winch:2024xdt, Torniamenti:2024uxl}.

Most observed BBHs are thought to contain two first-generation black holes formed from stellar collapse, creating a second-generation remnant~\cite{Zevin:2020gbd, Colloms:2025hib}.
In dense stellar environments, remnants may undergo subsequent coalescences, creating hierarchical mergers: those containing at least one component that is the product of one or more previous mergers~\cite{OLeary:2005vqo, Antonini:2016gqe, Rodriguez:2019huv, Doctor:2019ruh, Mahapatra:2024qsy, Sedda:2023qlx}. Hierarchical mergers are characterized by higher primary masses, isotropic spin orientations~\cite{OLeary:2005vqo}, and dimensionless spin magnitudes $\chi\approx0.7$, arising from the orbital angular momenta of their progenitors~\cite{Pretorius:2005gq, Buonanno:2007sv, Tichy:2008du, Rezzolla:2007rz, Hofmann:2016yih}, with some spread about $\chi\approx0.7$ due to their mass ratio and spins~\cite{Zevin:2022bfa, Borchers:2025sid}. These spin predictions are robust across binary star evolution assumptions~\cite{Fishbach:2017dwv}. Since hierarchical assembly produces more massive black holes, it is a natural candidate for filling the pair-instability supernova (PISN) gap.

To disentangle hierarchical mergers from first-generation counterparts, it is useful to look in parameters that are both well-measured and in which we expect distinct behavior. The rapid spins, unequal mass ratios, and/or high masses of several observed BBH signals---namely, GW190521~\cite{LIGOScientific:2020iuh}, GW231123~\cite{LIGOScientific:2025rsn} and GW241011~\cite{gw1011disc}---have led to their interpretation as hierarchical mergers~\cite{LIGOScientific:2020ufj, Kimball:2020qyd, Tagawa:2020qll, Gayathri:2021xwb, Gerosa:2021mno, Baibhav:2021qzw}, though nonhierarchical interpretations have also been put forward, e.g.~\cite{Fishbach:2020qag, Farrell:2020zju, Kinugawa:2020xws, Popa:2025dpz, Gottlieb:2025ugy, Croon:2025gol, Bartos:2025pkv, Tenorio:2026dcc}.
However, data quality issues and waveform inaccuracies can introduce systematic biases in measurements of individual systems, including their spins~\cite{gwtc3cat, LIGO:2021ppb, Capote:2024rmo, Payne:2022spz, Udall:2025bts, LIGOScientific:2025rsn, Ray:2025rtt, Tenorio:2026dcc}. When performing and interpreting population analyses, we therefore turn to \textit{effective parameters}, which are typically more reliably measured than component spin magnitudes and tilts~\cite{Macas:2023wiw, Miller:2024sui, Vitale:2022dpa}. Effective inspiral spin~\cite{Racine:2008qv, Gerosa:2017kvu} is typically the best-measured combination of spin parameters~\cite{Vitale:2016avz}, defined as the mass-weighted projection of the spin vectors onto the orbital angular momentum: $\chieff = (\chi_1\cos \tau_1+ q\chi_2\cos\tau_2)/(1+q)$; where $q$ denotes the mass ratio (defined on $0<q\leq1$), $\chi_1$ and $\chi_2$ are the respective spin magnitudes of the primary (heavier) and secondary (lighter) black holes, and $\tau_{1,2}$ are the angles between the spin vectors and the Newtonian orbital angular momentum at a chosen reference frequency.

The natal spins of black holes are theoretically uncertain but often expected to be small~\cite{Spruit:2001tz, Qin:2018vaa, Fuller:2019sxi, Fuller:2022ysb, Burrows:2023ffl}, in which case first-generation BBHs (1G+1G) will have a narrow $\chieff$ distribution centered near zero. Isotropically oriented dynamical mergers create a distribution symmetric around zero, while preferentially aligned isolated binaries would induce a positive centroid~\cite{Farr:2017uvj, Gerosa:2017kvu}. In contrast, hierarchical mergers between one first-generation and one second-generation black hole (1G+2G) will have a uniform $\chieff$ distribution centered at 0 with a width of $\sim 1$~\cite{Baibhav:2020xdf}. As higher-generation merger rates are likely an order of magnitude below that of 1G+2G mergers~\cite{Kremer:2019iul, Rodriguez:2021qhl, Ye:2024ypm}, we specialize to the latter.

As spin parameters are useful differentiators between formation channels, many studies have searched for population-level signatures in $\chieff$, spin magnitudes, and/or spin orientations, often in conjunction with mass dependence~\cite{Stevenson:2017dlk, Safarzadeh:2020mlb, Tiwari:2020otp, Hoy:2021rfv, Tagawa:2021ofj, Mould:2022ccw, Fishbach:2022lzq, Wang:2022gnx, Li:2023yyt, Hussain:2024qzl, Ray:2024hos, Adamcewicz:2025phm, Tiwari:2025oah, Berti:2025usa, Pierra:2024fbl, Banagiri:2025dmy, Antonini:2024het, Antonini:2025zzw, Tong:2025wpz, Antonini:2025ilj, Stegmann:2025zkb, Tong:2025xir, Farah:2026jlc, Vijaykumar:2026zjy}. Recently, Refs.~\cite{Pierra:2024fbl, Banagiri:2025dmy, Stegmann:2025zkb} found indication of a transition to higher spin magnitudes in the population at primary masses above $m_1\sim40\text{--}50M_\odot$, and Refs.~\cite{Antonini:2024het, Antonini:2025zzw, Tong:2025wpz, Antonini:2025ilj, Tong:2025xir} presented strong evidence for a transition in $\chieff$ from narrow to broad above $m_1\sim 45M_\odot$. References~\cite{Tong:2025wpz, Tong:2025xir} furthermore demonstrated the $\chieff$ transition to be coincident with a truncation in the secondary mass distribution, remarkably in line with the theoretical expectation for the lower edge of the PISN gap. These studies suggest that at $m_1\gtrsim 45M_\odot$ the population is dominated by second-generation black holes.

While effective spin has been a successful population-level diagnostic of hierarchical mergers, it is a lossy compression of the six-dimensional spin space to a single parameter. Hierarchical mergers are distinguished from 1G+1G mergers not only in $\chieff$, but cleanly in the \textit{joint} space of $\chieff$ and effective precessing spin $\chip$, which is a linearization of the in-plane spin dynamics~\cite{Schmidt:2014iyl, Gerosa:2020aiw, Thomas:2020uqj}: $\chip = \max[\chi_1\sin\tau_1,\ (3+4q)/(4+3q)\,q\chi_2\sin\tau_2]$. The joint $\chieff$--$\chip$ distribution for 1G+1G mergers is expected to cluster near $\chieff,\chip\sim0$~\cite{Payne:2024ywe}, while that of 1G+2G mergers scatters around the boundary of an ellipse centered on 0 with major axis $\chip \approx 0.7$ and minor axis $\chieff \approx 0.47$~\cite{Baibhav:2021qzw, Antonini:2024het}.

In this Letter, we introduce an astrophysically motivated population model in the joint space of effective inspiral and precessing spin. We include a two-dimensional ellipse-boundary component to capture 1G+2G mergers alongside an uncorrelated Gaussian to describe 1G+1G binaries. We explore the mixing fraction between these components as a function of mass and recover strong evidence for a subpopulation of hierarchical mergers at $m_\mathrm{peak} = 15.7_{-1.1}^{+3.2}M_\odot$\footnote{Here and elsewhere we quote the median and 90\% symmetric credible interval.} and a transition at  $m_* = \mstarcred M_\odot$ above which the majority of mergers are hierarchical. Our results are consistent with and bolster studies that model only in $\chieff$ by capturing the astrophysical expectation with higher fidelity in the multidimensional spin space. We reveal correlated structure in the $\chip$ distribution that increases credence both for the robustness of spin-mass features and their association with hierarchical mergers.

%%%%%%%%%%%%%%%%%%%%%%%%%%%%%%%%%%%%%%

% for my own readability
\hspace{0.1in}

\begin{figure}[!htbp]
\centering
\includegraphics[width=\columnwidth]{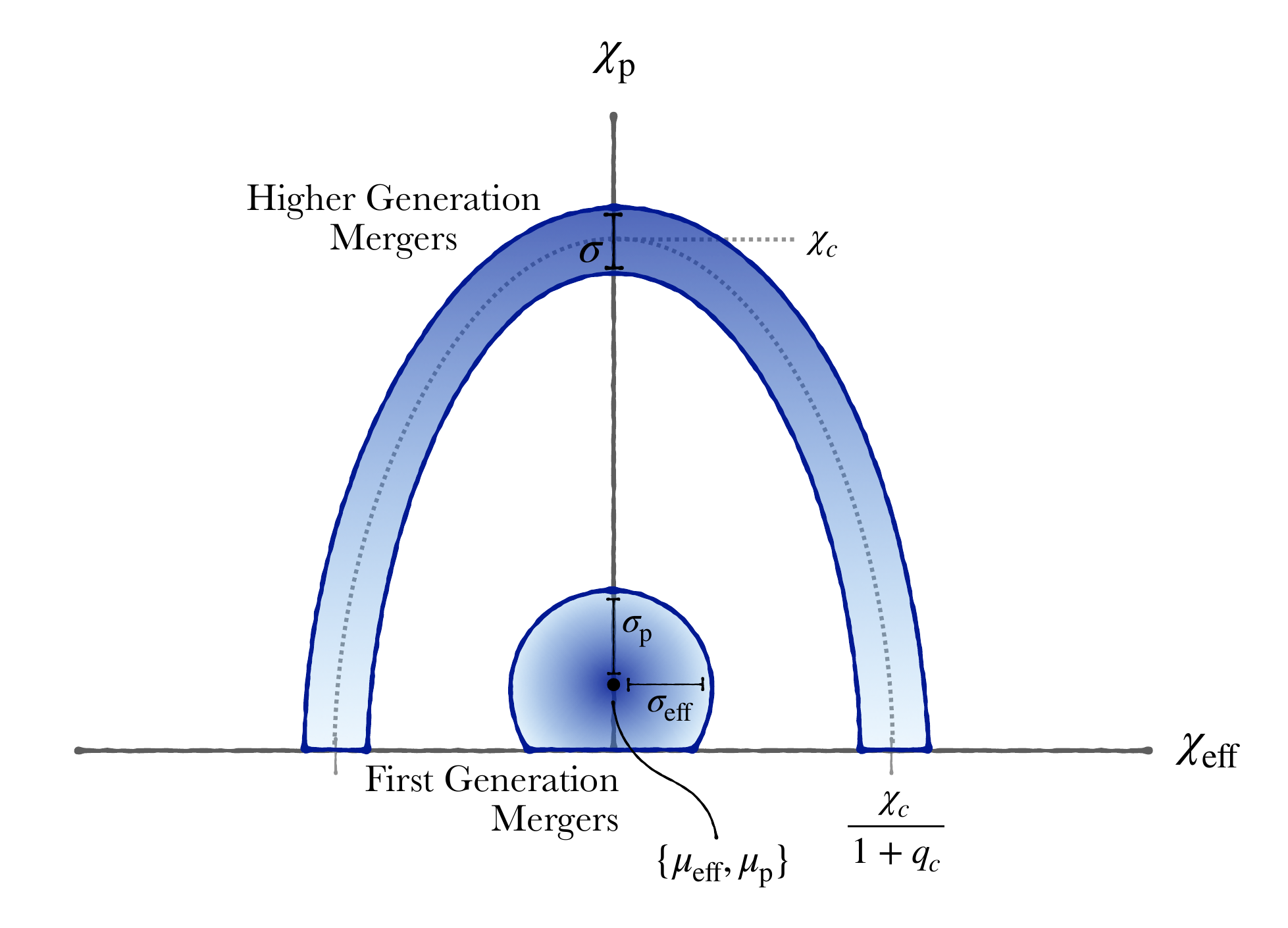}
\caption{Schematic of our two-component model in $\chieff$--$\chip$ space. The truncated Gaussian describes 1G+1G mergers and is defined by its means ($\mu_\mathrm{eff,p}$) and standard deviations ($\sigma_\mathrm{eff,p}$) in both dimensions. The ellipse-boundary component, ascribed to 1G+2G mergers, is determined by a characteristic spin $\chi_c$, characteristic mass ratio $q_c$, and Gaussian width perpendicular to the curve $\sigma$.}
\label{fig:model}
\end{figure}

\paragraph*{Models}---
\label{sec:models}
We analyze BBHs in the Fourth Gravitational-Wave Transient Catalog (GWTC-4)~\cite{gwtc4cat} with false-alarm rates $<1\,\mathrm{yr}^{-1}$ alongside GW241011 and GW241110, two events from the second part of the fourth observing run (O4b) whose spin properties imply a hierarchical origin~\cite{gw1011disc}. We model the population $\chieff$--$\chip$ distribution as a mixture of two components: independent truncated Gaussians for the 1G+1G distribution and a probability density concentrated on an ellipse boundary for the 1G+2G distribution.\footnote{Throughout, we associate the Gaussian and ellipse-boundary components to 1G+1G and 1G+2G mergers, respectively, and refer to the components with their associated astrophysical interpretations interchangeably.} Fig.~\ref{fig:model} schematically depicts the model components. The Supplemental Material contains further details of our models and hierarchical inference settings, including sensitivity injections and waveform models for single-event parameter estimation.

When secondary spin is negligible, $\chieff\approx \chi_1/(1+q)\cos\tau_1$ and $\chip\approx \chi_1\sin\tau_1$. Second-generation black holes have a characteristic spin $\chi_c\sim0.7$ and typical mass ratios $q_c\sim0.5$ with their secondary. With isotropic tilts, the astrophysical distribution of 1G+2G mergers in $\chieff$--$\chip$ follows the boundary of an ellipse with major axis $\chip=\chi_c\approx0.7$, minor axis $\chieff=\chi_c/(1+q_c) \approx 0.47$, uniform marginal $\chieff$, and marginal $\chip$ declining by $p(\chip)\propto\chip/(\chi_c^2-\chip^2)^{1/2}$~\cite{Baibhav:2020xdf, Baibhav:2021qzw, Antonini:2024het}. The separation in $\chieff$--$\chip$ between first-generation and hierarchical mergers, and the latter's elliptical structure, is pictured in Fig.~1 of Ref.~\cite{Payne:2024ywe} for simulated dynamical mergers from the \textsc{Cluster Monte Carlo} catalog~\cite{Kremer:2019iul, Rodriguez:2021qhl}.

We model the probability density for 1G+2G mergers as a Gaussian in the perpendicular distance from an ellipse boundary with a filter along $\chip$ to account for the declining density. The characteristic spin $\chi_c$, characteristic mass ratio $q_c$, and Gaussian width $\ln\sigma$ are inferred from the data. The $\chieff$ and $\chip$ distributions of the 1G+1G component are modeled as uncorrelated Gaussians, though we consider other models in Supplemental Material.
We allow the mixing fraction between the Gaussian (1G+1G) and ellipse-boundary (1G+2G) components to vary with primary mass, similar to the models employed in Refs.~\cite{Antonini:2025ilj, Tong:2025xir}. We use a cubic spline to flexibly model the branching ratio without imposing strong priors on the existence or location of transitions; Supplemental Material contains specifics of our node choices. While this is a strong model, we do not enforce the 1G+2G component's existence; further, it is built on simple, robust astrophysical grounds.

For primary mass, we make the simplifying assumption that the population is described by a single distribution, for which we use the default model of the LVK for GWTC-4: a broken power law with two Gaussian peaks and low-mass smoothing~\cite{gwtc4pop}. While we employ the standard power-law mass ratio model for the 1G+1G component, for the 1G+2G mergers we instead use a truncated normal, as hierarchical mergers may have a mass ratio peak away from unity. Evolution over cosmic time for both components is modeled with a single power law.

\hspace{0.1in}

\begin{figure}[!htbp]
\centering
\includegraphics[width=\columnwidth]{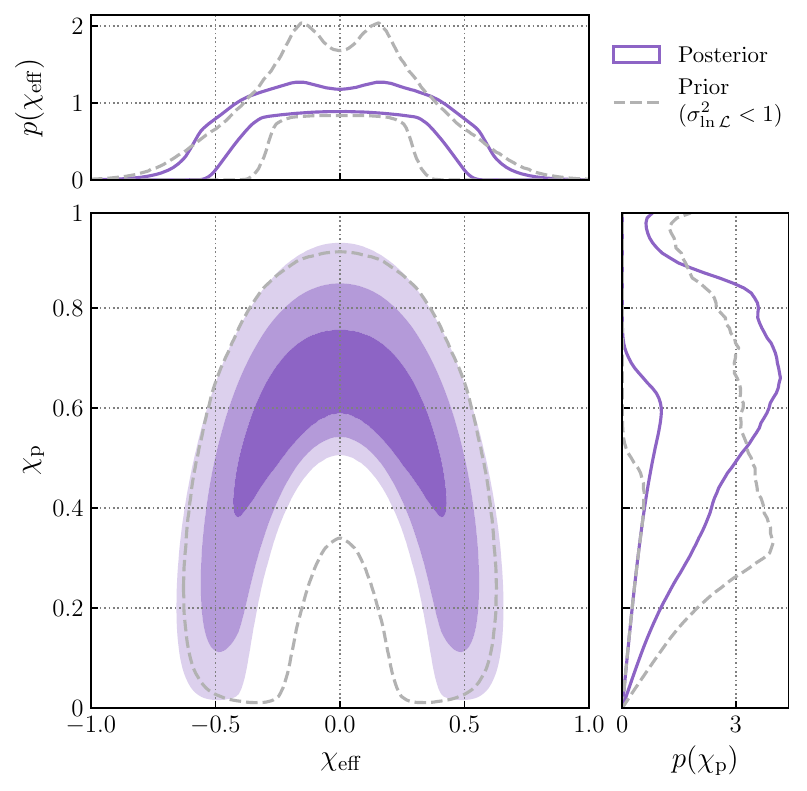}
\caption{Posterior on the ellipse-boundary component in $\chieff$--$\chip$ space. The joint distribution plots the posterior median with 50\%, 90\%, and 99\% percentile contours. In dashed gray, we show the 99\% credible contour of the median prior draw that passes the likelihood variance threshold. Side panels plot the 90\% credible regions of the marginal posteriors and priors.}
\label{fig:ellipse-posterior}
\end{figure}

\paragraph*{Results}---%
\label{sec:results}
Figure~\ref{fig:ellipse-posterior} shows the posterior on $\chieff$ and $\chip$ for the ellipse-boundary (1G+2G) component;  Supplemental Material Fig.~3 shows the full distribution with both components. The main panel shows the median joint distribution while the marginals show the 90\% credible regions of the posterior. Prior draws that pass the likelihood variance cut are in gray. The characteristic spin is inferred to be $\chi_c=0.66_{-0.28}^{+0.18}$, in agreement with the expectation of $\chi_c\approx0.7$ for 1G+2G mergers~\cite{Hofmann:2016yih}. Figure~\ref{fig:ellipse-posterior} illustrates the normalized density of the hierarchical component independent of its population fraction.

\begin{figure}[!htbp]
\centering
\includegraphics[width=\columnwidth]{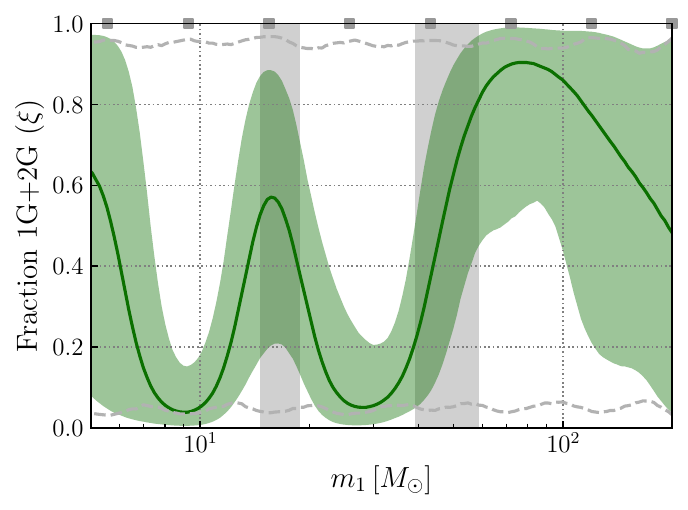}
\caption{The inferred fraction of hierarchical mergers $\xi(m_1)$. We plot the median and 90\% symmetric credible interval. Gray squares mark the locations of the spline nodes. The dashed lines enclose the 90\% region of the prior. The gray vertical regions span the 90\% credible regions for the identified low-mass peak $m_\mathrm{peak}$ and high-mass transition $m_*$.}
\label{fig:spline-posterior}
\end{figure}

Figure~\ref{fig:spline-posterior} plots the fraction of 1G+2G mergers, $\xi$, as a function of primary mass.
Near $m_1\sim46\,M_\odot$, the 1G+2G fraction transitions from $\xi\approx 0$ to $\xi>0$, consistent with Refs.~\cite{Pierra:2024fbl, Banagiri:2025dmy, Antonini:2024het, Antonini:2025zzw, Tong:2025xir, Tong:2025wpz, Antonini:2025ilj}. While the flexible model does not force a transition, we can use the inferred $\xi(m_1)$ structure to define a transition point $m_*$ as the mass in  $m_1\in(30,80)M_\odot$ at which $\xi$ first crosses 0.5. We find $m_* = \mstarcred M_\odot$. The 1G+2G fraction is confidently constrained to be nonzero above this transition---and is consistent with unity---until $\sim 100 M_\odot$. Quantitatively, we obtain $\xi>0.5$ for $m_1\in[60,100]M_\odot$ at 90\% credibility. Outside $\sim7\text{--}100M_\odot$, we roughly recover the prior. Marginalized over primary mass, the total fraction of hierarchical mergers is $\xi = 0.14_{-0.07}^{+0.12}$.

In addition to a high-mass population of hierarchical mergers, we observe evidence for a distinct subpopulation of 1G+2G mergers with masses $m_1\in(13,20) M_\odot$. The 1G+2G fraction is constrained to be $\gtrsim5\%$ at the 99\% level in this range. The existence of this low-mass population is compatible with Refs.~\cite{Antonini:2025ilj, Tong:2025xir} even as we expand the dimensionality and accuracy of the population model by including spin precession. Taking the value of $m_1$ that maximizes $\xi(m_1)$ in this mass range, we find a peak in the fraction of hierarchical mergers at $m_\mathrm{peak} = 15.7_{-1.1}^{+3.2}M_\odot$. Inference of this feature is not solely driven by the hierarchical candidates GW241011 and GW241110, which lie in this mass regime: an analysis excluding these events also recovers the lower peak (see Supplemental Material), indicating the feature is not an artifact of their inclusion. Including these events without the entirety of O4b introduces a degree of selection bias, though we note this choice is shared by other recent works~\cite{Tong:2025xir, Farah:2026jlc, Vijaykumar:2026zjy}, as GWTC-4 alongside these events comprise the most complete data product currently available. While the data suggest potentially a large fraction of mergers in 13--20 $M_\odot$ are hierarchical, they also imply few mergers in 20--40 $M_\odot$ are: the $\chieff$--$\chip$ distribution is inferred to be primarily Gaussian (i.e., first-generation) for $m_1\sim 8\text{--}10 \ M_\odot$ and $m_1\sim 20\text{--}40 \ M_\odot$. Our tightest constraints on $\xi$ are in the regimes where we infer a low proportion of hierarchical mergers. At $m_1=9M_\odot \ (30M_\odot)$, we constrain $\xi<0.12 \ (\xi<0.16)$ at 90\% credibility.

The astrophysically informed ellipse-boundary $\chieff$--$\chip$ model for the hierarchical component induces, by design, a quasiuniform marginal $\chieff$ akin to the one-dimensional model employed by Refs.~\cite{Tong:2025xir, Antonini:2025ilj}. Hence, our comparable identification of spin-mass features is not unexpected. The inclusion of $\chip$ does not significantly improve constraints (see Supplemental Material), but crucially corroborates the hierarchical hypothesis in the precessing spin dimension. Further, $\chip$ enables richer inference of the population spin structure, such as the inferred spin magnitude peak at $\chi\approx0.7$, which strengthens the emerging picture of a high-spin, isotropic subpopulation consistent with hierarchical assembly.

The inclusion of the 1G+2G population significantly impacts the inferred spin distribution of the bulk BBH population (also see Supplemental Material Fig.~3). The width of the 1G+1G Gaussian is, foreseeably, narrower than when the hierarchical component is excluded. In $\chieff$, the two-component model recovers $\sigma_{\chieff}=0.07_{-0.01}^{+0.02}$, nearly half that of the Gaussian-only model, which returns $\sigma_{\chieff}=0.13_{-0.02}^{+0.03}$. The $\chieff$ mean remains constrained positive, $\mu_{\chieff} = 0.03_{-0.02}^{+0.02}$. Neglecting the 1G+2G subpopulation biases first-generation spin measurements as a one-component model artificially broadens to absorb the hierarchical spread.

The model including the hierarchical component, with the branching ratio an evolving function of mass, is preferred over a pure Gaussian model with a natural log Bayes factor of $\ln \mathcal B = 14.5$, which accounts for model complexity. It is favored over a model where the branching ratio is constant over mass with $\ln\mathcal B = 6.8$. These Bayes factors indicate significant evidence for structure in the effective inspiral and precessing spin distribution consistent with hierarchical mergers, which, moreover, is nonlinearly correlated with primary mass.

\hspace{0.1in}

\paragraph*{Discussion and conclusion}---
\label{sec:conclusion}
Using an astrophysically informed mixture model in the joint $\chieff$--$\chip$ space, we find strong evidence for a subpopulation of binary black hole mergers that occupy the boundary of an ellipse with a characteristic spin near $\chi_c\sim0.7$, consistent with the expectation for hierarchical mergers. Allowing the mixing fraction between the components to vary with primary mass, we identify a transition to primarily high-spin and quasiuniform $\chieff$ at $m_* = \mstarcred M_\odot$. Interpreting the elliptical arc component as hierarchical mergers, the data indicate most black holes above $46M_\odot$ arise from repeated mergers in dense stellar clusters, aligned with the predicted onset of the PISN gap.

Moreover, we measure a peak in the fraction of hierarchical mergers at $m_\mathrm{peak} = 15.7_{-1.1}^{+3.2}M_\odot$. Hierarchical assembly is frequently discussed as a method to produce black holes in and above the PISN gap. Indeed, the transition near $46 M_\odot$ corroborates this interpretation. However, simulations of dense stellar clusters suggest the 2G mass distribution declines with primary mass and peaks near $15M_\odot$, driven by an abundance of low-mass first-generation black holes produced by solar-metallicity stars~\cite{Rodriguez:2019huv, Kremer:2019iul, Ye:2024ypm, Ye:2025ano}.  As such, the observation of this lower peak in BBH data is unsurprising and may constitute the maximum of the 2G mass spectrum. At the same time, 2G black holes above $46 M_\odot$ are uncommon in near-solar-metallicity clusters, potentially implying the high-mass component arises from clusters with different initial conditions---namely, lower metallicities---than the low-mass population. Coupling the features in the data with predictions from stellar evolution may enable constraints on the metallicity distribution of the progenitor stars of black-hole mergers.

While the low-mass peak meshes with theoretical expectation, the ostensible \textit{lack} of hierarchical mergers between 20 and 40 $M_\odot$ is perhaps more surprising. The dominance of hierarchical mergers in the $10\text{--}20 \ M_\odot$ range followed by the dearth in $20\text{--}40 \ M_\odot$ may imply either a dip in the first-generation mass spectrum in the former range, or a pileup in the latter. A drop in progenitor core compactness may drive a dip in this lower regime~\cite{Schneider:2020vvh}, for which several studies have searched and some have tentatively identified~\cite{Galaudage:2024meo, Adamcewicz:2024jkr, Willcox:2025poh}. If the observed feature at $\sim35M_\odot$ primarily  contains first-generation mergers, it may also contribute to the apparent lack of second-generation black holes at moderate masses. Untangling what the inferred branching fraction across primary mass indicates for both the first- and second-generation mass spectra---and the properties of their stellar progenitors---is a key avenue for future study.

Although past work has yielded similar results, no studies have thus far considered the influence of precessing degrees of freedom. Neglecting to model a parameter (which, in practice, means assuming it follows a fixed distribution---here, the single-event parameter estimation prior) can be a source of systematic bias. While we find $\chip$ does not substantially further constrain the mass-dependent fraction of hierarchical mergers obtained with $\chieff$ only, the consistency between results bolsters confidence in the hierarchical interpretation of effective spin structure while extending into a deeper observable space.

Taken at face value, the fraction of hierarchical mergers inferred from GW data is consistent with a population of entirely dynamical mergers. This meshes with Ref.~\cite{Rodriguez:2021nwd}, who note consistency between the BBH merger rate found by the LVK and from cluster simulations. We find the total 1G+2G fraction to be $\xi\sim14\%$, while 1G+2G mergers comprise $\sim17\%$ of mergers in the \textsc{Cluster Monte Carlo} catalog~\cite{Kremer:2019iul, Rodriguez:2021qhl}, and the astrophysical hierarchical fraction is likely closer to $\sim10\%$~\cite{Ye:2025ano}. The 1G+1G $\chieff$ distribution has support at negative values, which may signify a large contribution of 1G dynamical mergers, as expected given the apparent observation of 2G dynamical mergers. However, its confidently positive centroid is instead indicative of preferentially aligned isolated binaries. We caution against overinterpreting the similarity in population fraction, as both rely on uncertain assumptions: the former on associating the ellipse-boundary component to 1G+2G mergers, and the latter on prescriptions that control the merger rate such as the escape velocity distribution.

Sequential mergers may also occur in active galactic nuclei disks, preferentially creating aligned mergers~\cite{Yang:2019cbr, Tagawa:2020qll, Vaccaro:2023cwr, Li:2025iux}. A symmetric ellipse arc is favored (see Supplemental Material), suggesting active-galactic-nuclei-driven mergers are not dominant constituents of the inferred $\chieff$--$\chip$ distribution. Further observations might allow us to elucidate hierarchical subpopulations corresponding to different mechanisms that prompt the repeated mergers.

Linking the black-hole mass spectrum to features in the spin distribution is a promising avenue to more precisely identify hallmarks of stellar evolution that may otherwise be washed out, like the PISN mass gap, by combining information across multiple dimensions. Improved constraints on the PISN gap will sharpen bounds on nuclear processes inaccessible at laboratory temperatures, in particular the cross section of the $^{12}\mathrm{C}(\alpha,\gamma)^{16}\mathrm{O}$ reaction, pivotal to energy production in massive stars~\cite{deBoer:2017ldl, Farmer:2020xne, Costa:2020xbc}. As the catalog of binary black hole mergers continues to grow alongside advancements in stellar theory and simulation, the statistical significance and associated astrophysical interpretation of spin-mass features in the population will be tested with increasing rigor, emphasizing the power of astrophysically informed population studies to uncover the evolutionary history of black holes.

\hspace{0.1in}

\paragraph*{Acknowledgments}---
\label{sec:acknowledgments}
We thank Sofía Álvarez-López, Fabio Antonini, Sharan Banagiri, Thomas Dent, Maya Fishbach, Jack Heinzel, Asad Hussain, Matthew Mould, Gregoire Pierra, Vaibhav Tiwari, Hui Tong, and Noah Wolfe for discussions.
C.P. is supported by the National Science Foundation Graduate Research Fellowship under grant no. DGE-2141064.
M.Z. gratefully acknowledges funding from the Brinson Foundation in support of astrophysics research at the Adler Planetarium. 
S.V. is partially supported by the NSF grant no. PHY-2045740.

The authors are grateful for computational resources provided by the LIGO Laboratory supported by National Science Foundation grants no. PHY-0757058 and no. PHY-0823459.
This material is based upon work supported by NSF's LIGO Laboratory which is a major facility fully funded by the National Science Foundation and has made use of data or software obtained from the Gravitational Wave Open Science Center (gwosc.org), a service of the LIGO Scientific Collaboration, the Virgo Collaboration, and KAGRA.

\paragraph*{Data availability}--- The data that support the findings of this article are openly available at Ref.~\cite{datarelease}.

\bibliography{paper}

\clearpage

\section{Supplemental Material}
\subsection{Model details}\label{app:models}

The relation between $\chieff$ and $\chip$ when secondary spin is small is well-described by
\begin{align}
    \chip^2 = a_1^2-(1+q)^2\chieff^2.
\end{align} This defines the boundary of an ellipse,
\begin{align}
    F(\chieff,\chip|\chi_c,q_c)=\left(\frac{1+q_c}{\chi_c}\right)^2\chieff^2+\frac{\chip^2}{\chi_c^2}-1 = 0,
\end{align} where $\chi_c$ and $q_c$ are the characteristic primary spin and mass ratio. The distribution has some spread around this curve due to variation in mass ratio, primary spin, and secondary spin. To capture this width, we model the probability density for this component as a Gaussian in the distance to the curve. The perpendicular distance from a point $(\chieff,\chip)$ to $F$ is, to linear order
\begin{align}
    d(\chieff,\chip|\chi_c,q_c)&\approx\frac{|F(\chieff,\chip|\chi_c,q_c)|}{\Vert\nabla F(\chieff,\chip|\chi_c,q_c)\Vert}\\
    &= \frac{\left|\frac{\chieff^2}{(\chi_c/(1+q_c))^2}+\frac{\chip^2}{\chi_c^2}-1\right|}{2\left(\frac{\chieff^2}{(\chi_c/(1+q_c))^4}+\frac{\chip^2}{\chi_c^4}\right)^{1/2}}.
\end{align} The density at $(\chieff,\chip)$ is thus
\begin{align}
    \mathcal N(\chieff,\chip|\chi_c, q_c, \sigma) \propto\exp\left(-\frac{d(\chieff,\chip|\chi_c,q_c)^2}{2\sigma^2}\right).
\end{align} An isotropic tilt angle distribution, as expected for hierarchical mergers, corresponds to a $\sin\tau_1$ distribution  diverging at unity, $p(\sin\tau_1)\propto\sin\tau_1/\sqrt{1-\sin^2\tau_1}$. The marginal distribution of $\chip$ thus peaks at $\chi_c$ following
\begin{align}
    p(\chi_p)\propto\begin{cases}
        \frac{\chip}{\sqrt{\chi_c^2-\chip^2}},& \chi_p<\chi_c;\\
        0, & \chi_p\ge\chi_c.
    \end{cases}
\end{align} As divergent behavior and sharp cutoffs are problematic to model, we instead accommodate this falloff with the filter $f(\chip)=\chip/\sqrt{1-\chip^2}$. While this approximation lacks steepness toward $\chi_c$, we find the Gaussian overlay accommodates the necessary density toward $\chi_c$ such that this model can reproduce the marginal $\chip$ distribution of simulated 1G+2G mergers~\cite{Kremer:2019iul, Rodriguez:2019huv} quite accurately.
The total probability density for the ellipse component is
\begin{align}
    p_\mathrm{1G+2G}(\chieff,\chip|\chi_c,q_c,\sigma) = \mathcal N(\chieff,\chip|\chi_c, q_c, \sigma) f(\chip).
\end{align} 
The width parameter $\sigma$ captures the expected spread in the $\chieff$--$\chip$ distribution due to the varying mass ratios and precise spin values of 2G mergers. We fit for the width and characteristic spin and mass ratio values with priors listed in Tab.~\ref{tab:priors}.

\subsection{Hierarchical Bayesian inference}\label{app:inference}

We sample the population posterior using \texttt{Dynesty}~\cite{Speagle:2019ivv} and \texttt{Bilby}~\cite{bilby_paper} with the population likelihood~\cite{Mandel:2018mve, Vitale:2020aaz} as implemented in \texttt{GWPopulation}~\cite{Talbot:2024yqw}. Our catalog consists of the 153 binary black holes in GWTC-4~\cite{gwtc4cat} and the special events GW241011 and GW241110~\cite{gw1011disc}. We use the \textsc{Mixed} waveform results in the publicly-released posterior samples~\cite{gwtc4release, gw1011release} for most events, but reduce the parameter estimation contribution to the log-likelihood variance by instead using the \textsc{IMRPhenomXPHM}~\cite{Pratten:2020ceb} results for GW150914 and GW200129, which have fewer than 5000 \textsc{Mixed} samples. We estimate survey sensitivity with the public set of BBH injections recovered by search pipelines spanning all four observing runs~\cite{Essick:2025zed}.

We limit the log-likelihood variance from the Monte Carlo approximations in the likelihood to be smaller than 1, a condition shown to be sufficient to prevent bias in the posterior~\cite{Essick:2022ojx, Heinzel:2025ogf}. We compute the error statistic introduced in Ref.~\cite{Heinzel:2025ogf} and find an information loss of 0.18 bits, within the suggested threshold. To test the impact of the variance cut, we also run with a limit of 4. The primary difference is in the inferred 1G+1G $\chip$ distribution, which is narrower and closer to zero than found with a variance cut of 1. While the data seemingly prefer small $\chip$ values, our inferential capability is limited in that regime predominantly due to selection variance.

\begin{table}
\centering
\begin{tabular}{ccc}
\hline \hline
Symbol & Description & Prior  \\
\hline \hline
$\xi$ & Fraction of 1G+2G black holes & $\mathcal U(0,1)$ \\
\hline
$f_i$ & $i^\mathrm{th}$ spline node latent value & $\mathcal N(0,2)$ \\
\hline
$\lambda$ & redshift power-law index & $\mathcal U(0,10)$ \\
\hline
$\alpha_1$ & $m_1$ power-law index below break & $\mathcal U(-4,12)$ \\
$\alpha_2$ & $m_1$ power-law index above break & $\mathcal U(-4,12)$ \\
$b_m$ & $m_1$ power-law break mass & $\mathcal U(20,50)$ \\
$\mu_1$ & Lower $m_1$ peak mean $[M_\odot]$ & $\mathcal U(5,20)$ \\
$\mu_2$ & Higher $m_1$ peak mean $[M_\odot]$ & $\mathcal U(25,60)$ \\
$\sigma_1$ & Lower $m_1$ peak width $[M_\odot]$ & $\mathcal U(0,10)$ \\
$\sigma_2$ & Higher $m_1$ peak width $[M_\odot]$ & $\mathcal U(0,10)$ \\
$\lambda_0$ & $m_1$ power law fraction & Dirichlet \\
$\lambda_1$ & $m_1$ lower peak fraction & Dirichlet \\
$\lambda_2$ & $m_1$ higher peak fraction & Dirichlet \\
$\delta_\mathrm{min,1}$ & $m_1$ low-mass smoothing width $[M_\odot]$ & $\mathcal U(0,10)$ \\
$\delta_\mathrm{min,2}$ & $m_2$ low-mass smoothing width $[M_\odot]$ & $\mathcal U(0,10)$ \\
$m_\mathrm{min,1}$ & $m_1$ minimum mass $[M_\odot]$ & $\mathcal U(3,10)$ \\
$m_\mathrm{min,2}$ & $m_2$ minimum mass $[M_\odot]$ & $\mathcal U(3,10)$ \\
$m_\mathrm{max}$ & Maximum mass $[M_\odot]$ & $=200$ \\
$\beta$ &  1G+1G $q$ power-law index & $\mathcal U(-2,7)$ \\
$\mu_q$ &  1G+2G $q$ mean & $\mathcal U(0,1)$ \\
$\sigma_q$ &  1G+2G $q$ width & $\mathcal U(0.02,1)$ \\
\hline
$\mu_{\chieff}$ & 1G+1G $\chieff$ mean & $\mathcal U(-0.3,0.3)$ \\
$\sigma_{\chieff}$ & 1G+1G $\chieff$ width & $\mathcal U(0.02,0.3)$ \\
$\mu_{\chip}$ & 1G+1G $\chip$ mean & $\mathcal U(0,0.5)$ \\
$\sigma_{\chip}$ & 1G+1G $\chip$ width & $\mathcal U(0.02,0.4)$ \\
\hline
$\chi_c$ & 1G+2G characteristic spin & $\mathcal U(0.3,1)$ \\
$q_c$ & 1G+2G characteristic mass ratio & $\mathcal U(0, 1)$ \\
$\ln\sigma$ & 1G+2G ellipse width & $\mathcal U(-4,-1)$ \\
\hline \hline
\end{tabular}
\caption{Priors on the model parameters, separated by model component. For the branching fraction, we use either the constant $\xi$ or the spline $f_i$.}
\label{tab:priors}
\end{table}

For the primary mass model, we adopt the standard priors used by the LVK~\cite{gwtc4pop}. To improve prior draw efficiency, we limit the prior on the redshift power-law index to $\mathcal U(0,10)$ as the entire posterior is contained above zero.

We model the spline using 10 equally-log-spaced nodes between $m_1 = 2M_\odot$ and $m_1=200 M_\odot$. Other numbers of nodes produce no tangible differences in results and 10 nodes has the best evidence. The latent values of the spline nodes are sampled from independent normal distributions $\mathcal N(0,2)$ and are transformed to the unit interval with a sigmoid. We limit the overlap between the $\mu_{\chip}$ and $\chi_c$ priors to prevent degeneracy in the likelihood, as the population is dominated by the lower-mean Gaussian component.
All priors are given in Tab.~\ref{tab:priors}.

\subsection{Additional results and model variations}\label{app:variations}

To directly assess the contribution of $\chip$, we perform an analysis in which $\chip$ is modeled with a single component, uncorrelated with a two-component (uniform and Gaussian) $\chieff$ model. The Bayes factor between the correlated and uncorrelated models is indistinguishable; however, Bayes factors are sensitive to prior volumes and should be interpreted with caution. The correlated model yields mildly reduced uncertainties on $\xi(m_1)$. More importantly, the consistency between the correlated model and both the uncorrelated and $\chieff$-only analyses is a meaningful robustness check: the data could have preferred $\chip$ outside the predicted elliptic region, contradicting the hierarchical scenario, but instead demonstrate full consistency with the theoretical expectation.

\begin{figure*}[!htbp]
\centering
\includegraphics[width=0.9\textwidth]{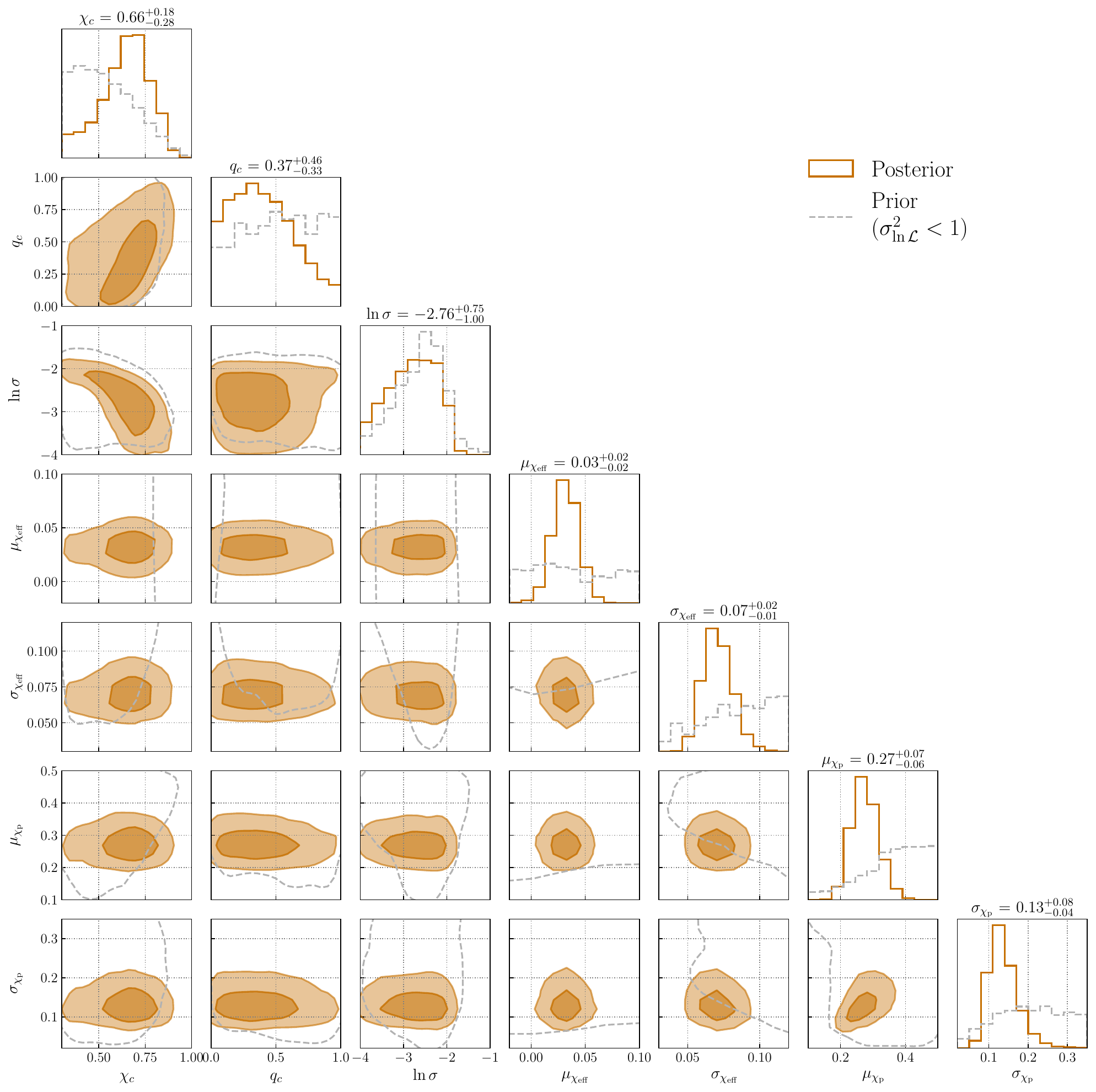}
\caption{Marginal and two-dimensional joint posteriors on the ellipse-boundary and Gaussian spin hyperparameters. The joint contours enclose 50\% and 90\% of the posterior volume. The 90\% contour and marginals of draws from the prior that pass the likelihood variance cut are shown in dashed gray.}
\label{fig:ellipse-corner}
\end{figure*}

The posteriors on the spin hyperparameters are in Fig.~\ref{fig:ellipse-corner}. The positive correlation between the characteristic spin $\chi_c$ and mass ratio $q_c$ arises from their inverse effects on the width of the marginal $\chieff$ distribution. The inference on the width $\ln\sigma$ is primarily driven by the variance cut, but with slight preference for narrower spread. We confirm through a leave-one-out analysis that these results are not driven solely by the exceptional event GW231123, which may possess near-extremal spins. Including GW231123 narrows the ellipse in $\chieff$ (increases $q_c$) and raises it in $\chip$ (increases $\chi_c$), but results are consistent to well within uncertainty.

\begin{figure}[!htbp]
\centering
\includegraphics[width=\columnwidth]{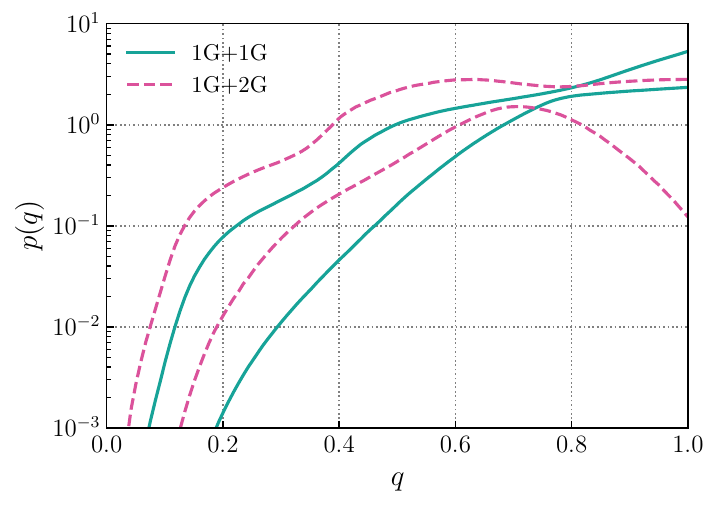}
\caption{The 90\% credible bounds on the marginal mass ratio distributions for the first-generation component (solid teal), modeled by a truncated power law, and the hierarchical component (dashed pink), modeled with a truncated normal.}
\label{fig:pq}
\end{figure}

Theoretical models for the 1G+2G mass ratio distribution find a peak near $q=0.5$~\cite{Kremer:2019iul, Rodriguez:2021qhl}; as such, we separately model $p(q)$ for 1G+1G and 1G+2G with results in Fig.~\ref{fig:pq}, where 1G+1G is in solid teal and 1G+2G in dashed pink. While the 1G+2G result is shifted away from unity vis-à-vis the 1G+1G distribution, the hyperparameters $\mu_q$ and $\sigma_q$ are weakly constrained, indicating the data are not strongly informative about the mass ratio distribution of the hierarchical component. We test the impact of the mass ratio model by contrasting analyses using a single power law, independent power laws, and independent Gaussians. All three alternate models are disfavored by $\ln\mathcal{B}\sim 1$ and the inferred spin distributions are nearly identical, suggesting that of the models explored, the chosen power law and Gaussian mixture best describes the mass ratio data.

\begin{figure}[!htbp]
\centering
\includegraphics[width=\columnwidth]{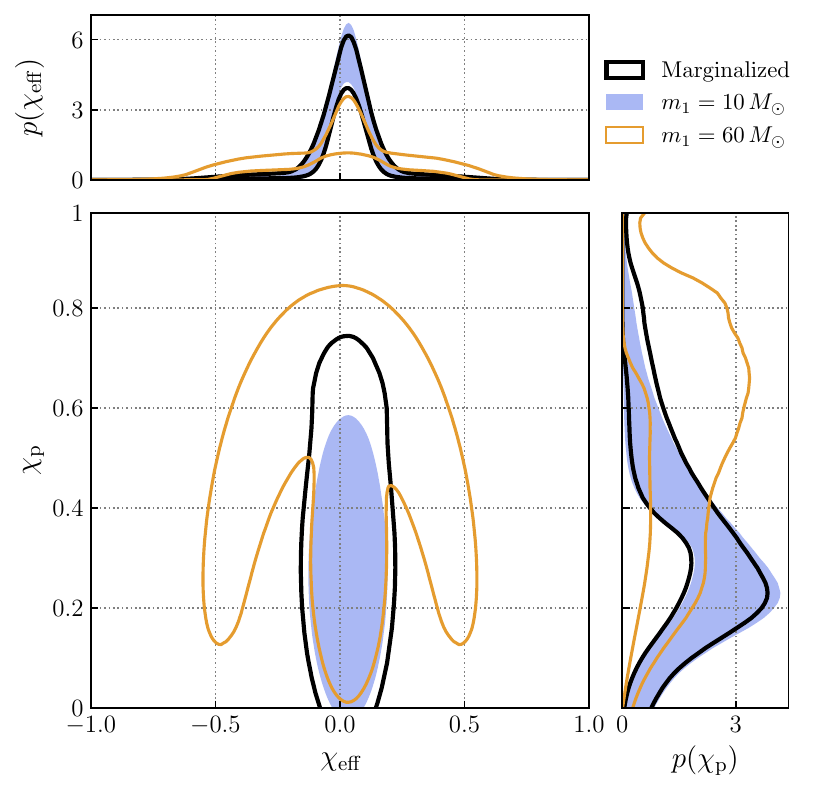}
\caption{The joint and marginal distributions of the total spin distribution marginalized over mass (black), at $m_1=10\,M_\odot$ (filled blue), and at $m_1=60\,M_\odot$ (gold). The contours enclose the 90\% boundary of the median while the marginals plot the 90\% credible region.}
\label{fig:ellipse-slices}
\end{figure}

Fig.~\ref{fig:ellipse-slices} shows the total $p(\chieff,\chip)$, both at slices and marginalized over mass:
\begin{align*}
    p(\chieff,\chip|m_1,\Lambda) &= (1-\xi(m_1|\Lambda))p_\mathrm{1G}(\chieff,\chip|\Lambda)\\
    &+\xi(m_1|\Lambda)p_\mathrm{2G}(\chieff,\chip|\Lambda);\\
    p(\chieff,\chip|\Lambda) &= \int\dd m_1 \, p(\chieff,\chip|m_1,\Lambda) \, p(m_1|\Lambda).
\end{align*} At $m_1=10\,M_\odot$, the population is found to be primarily 1G+1G, resulting in a narrow $p(\chieff)$ and $p(\chip)$ peaked at low values. In contrast, at $m_1=60\,M_\odot$, we infer a high fraction of hierarchical mergers, so the ellipse-arc component is visible in the joint panel and the marginals reflect the resulting broad $p(\chieff)$ and extended $p(\chip)$.

\begin{figure}[!htbp]
\centering
\includegraphics[width=\columnwidth]{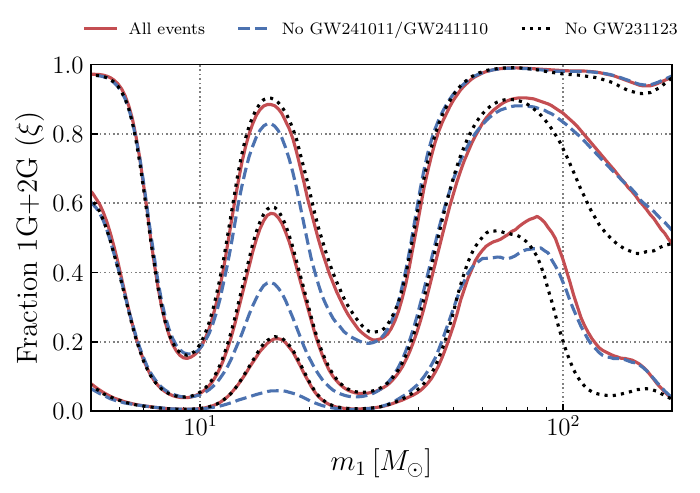}
\caption{The fraction of hierarchical mergers $\xi(m_1)$. In red, we analyze all 153 events in GWTC-4 alongside GW241011 and GW241011. In dashed blue, we only analyze the 153 GWTC-4 events. In dotted black, we include GW241011 and GW241110 but exclude GW231123.}
\label{fig:posterior-spline-exclude}
\end{figure}

The inferred low-mass subpopulation does not entirely depend on GW241011 and GW241110. Removing these events, the peak remains centered at $m_\mathrm{peak}=15.7M_\odot$, but the inferred fraction of hierarchical mergers at $m_\mathrm{peak}$ is reduced from $\xi(m_\mathrm{peak}) = 0.57_{-0.37}^{+0.31}$ to $\xi(m_\mathrm{peak}) = 0.37_{-0.31}^{+0.46}$. In Fig.~\ref{fig:posterior-spline-exclude} we show the inferred $\xi(m_1)$ using all 155 events (solid red), excluding GW241011 and GW241110 (dashed blue), and excluding GW231123 (dotted black). Evidence for the low-mass subpopulation exists in the GWTC-4 catalog alone; the selected O4b events add strength to the interpretation but do not create it.

We also test skew-normal and correlated models for the 1G+1G population, as recently employed by the LVK~\cite{gwtc3pop, Banagiri:2025dxo, gwtc4pop}, and find both are disfavored over an uncorrelated Gaussian 1G+1G component. The posteriors on the skewness and correlation coefficient are centered near and consistent with zero. This suggests the inferred skewness in Ref.~\cite{gwtc4pop} comes from a need to accommodate a dominant population with primarily positive $\chieff$ and a subdominant group that includes negative $\chieff$. However, a two-component model with a symmetric, narrow dominant component better fits the data. Finally, we relax the assumption of symmetry in the ellipse-arc component about $\chieff=0$ by fitting for its center. The posterior on the centering parameter is consistent with zero and the non-centered model is disfavored.

\subsection{Individual-event Bayes factors}\label{app:event-bf}

We compute the Bayes factor for each event to come from the 1G+2G over the 1G+1G component both for the constant $\xi$ and evolving $\xi(m_1)$ models as
\begin{align}
    \mathcal B_{1G+1G}^{1G+2G}(n) &= \frac{p(G_n = 1G+2G | \{d\})}{p(G_n = 1G+1G | \{d\})},
\end{align} where $G_n$ is the generation of the $n^\mathrm{th}$ event, and the marginal posterior on generation is
\begin{align}
    p(G_n| \{d\}) = &\int\dd\Lambda\dd\theta_n\, p(\Lambda|\{d\}) \, p(G_n| \Lambda)\nonumber\\
    &\times p(\theta_n|G_n,\Lambda) \frac{p(\theta_n|d_n)}{p(\theta_n)}\frac{p(d_n)}{p(d_n|\Lambda)}.
\end{align} The normalization term $p(d_n|\Lambda)/p(d_n) = \int\dd\theta\,p(\theta|d_n)p(\theta|\Lambda)/p(\theta)$. In practice, these integrals are computed with Monte Carlo sums over individual-event samples $\theta$ and hyperposterior samples $\Lambda$.

\begin{table}[h]
\centering
\begin{tabular}{lcc}
\hline\hline
Event & $\ln\mathcal{B}$ (const.\ $\xi$) & $\ln\mathcal{B}$ (evolving $\xi(m_1)$) \\
\hline
GW241011        & 7.6  & 10.5 \\
GW190517        & 4.1  & 3.6  \\
GW231028\_153006 & 3.2 & 6.3  \\
GW241110        & 2.5  & 3.6  \\
\hline\hline
\end{tabular}
\caption{Events with the highest Bayes factors to come from the 1G+2G component under the constant $\xi$ and evolving $\xi(m_1)$ models.}
\label{tab:bayes_factors}
\end{table}

Table~\ref{tab:bayes_factors} lists the events with the highest Bayes factors for being in the 1G+2G component for both the constant and evolving $\xi$ models. While the top four events are shared between the models, GW231028\_153006 and GW190517 swap ranks under the evolving model, as the former's well-constrained $m_1\gtrsim 60 M_\odot$ boosts its significance under the mass-dependent branching fraction. GW231123 ranks sixth using a constant $\xi$ and eighth under the evolving model; its lower significance in the latter may reflect the broadening of $\xi(m_1)$ toward the prior at primary masses as high as $\sim120\,M_\odot$. Several of these candidates overlap with those identified in other works~\cite[e.g.][]{Kimball:2020qyd, Tong:2025xir}. In a companion paper~\cite{Plunkett:2026inprep}, we use a machine-learning-based cluster simulation emulator to infer the host cluster properties of the events with the highest Bayes factors to come from the 1G+2G component, as well as the events with the highest $\chip$ likelihood ratios, a classification metric introduced by Ref.~\cite{Payne:2024ywe}.

%\bibliography{paper}

% \end{document}

\end{document}